# Coupling broadband terahertz dipoles to microscale resonators

Authors:

Christopher Rathje[1], Rieke von Seggern[1], Leon A. Gräper[1], Jana Kredl[2], Jakob Walowski[2], Markus Münzenberg[2] and Sascha Schäfer[1*]

Affiliations:

[1]Institute of Physics, University of Oldenburg, 26129 Oldenburg, Germany.
[2]Institute of Physics, University of Greifswald, 17489 Greifswald, Germany.

[*]Corresponding author: sascha.schaefer@uni-oldenburg.de

## Abstract

Spintronic emitters are a unique class of terahertz (THz) sources due to their quasi-two-dimensional geometry and thereby their capability to couple to resonator near fields. Global excitation of the emitters often obstructs the intricate details of the coupling mechanisms between local THz dipoles and the individual modes of resonator structures. Here, we demonstrate the spatial mapping of the coupling strength between a local terahertz source on a spintronic emitter and far-field light mediated by a structured metallic environment. For a bow-tie geometry, experimental results are reproduced by a numerical model providing insights into the microscopic coupling mechanisms. The broad applicability of the technique is showcased by extracting the THz mode structure in split-ring resonator metasurfaces and linear arrays. With these developments, planar THz sources with tailored spectral and angular emission profiles are accessible.

## Introduction

The introduction of spintronic terahertz emitters (STE)[1,2] consisting of stacks of ultrathin ferromagnetic and non-magnetic layers opened up a new class of pulsed terahertz (THz) sources, exhibiting ultrabroad and gapless emission spectra. In these emitters, a spin-polarized current is launched by femtosecond optical excitation within the magnetic layer and is subsequently deflected by the inverse spin Hall effect in a transverse direction orthogonal to the local spin polarization. The generated transient electric dipole emits broadband THz radiation into the far field.

In terms of THz emission efficiency, many improvements over the original design have been achieved in recent years[3–7], optimizing the layer configuration[8,9], film thicknesses[2,9–14], material compositions and crystallinity[2,11,12,15] and thin film interfaces[16–18]. The dependence of the inverse spin Hall deflection on the local magnetization enables the control of the THz polarization state and even polarization patterns by tuning the magnetization profile of the emitter[19–23]. Similarly, in-situ tuning of the thin-film strain induced by an adjacent biased ferroelectric material was demonstrated modifying the emitted THz field strength and the shape of the THz transient[24,25]. In other applications, the generated transverse current pulse is directly utilized and, for example, coupled into planar waveguides[26].

Furthermore, the quasi-two-dimensional geometry of the emitter allows for a direct emitter patterning[9,22,23,27–31] or a coupling of the emitter surface to metallic microstructures[32,33] and metasurfaces[34,35], aiming for a spectral and angular tuning of the emitted radiation, an improved excitation[36] or outcoupling efficiency[37], or enhanced sensitivity in local THz spectroscopy[38]. Up to now, spectral tunability by employing two-dimensional resonators and antennas combined with a large-area illumination of the spintronic emitter only yielded moderate effects.

As a pre-requisite for the successful design of patterned STEs, a microscopic understanding of the coupling of local THz dipoles to a photonic environment is essential. On the experimental side, several methodologies have been developed to map local material response in the THz frequency range, such as laser-induced THz emission microscopy[39–42] and THz- and mid-infrared variants of scanning near-field optical microscopy (SNOM)[43–46]. These, however, were not yet applied for characterizing the THz photonic environment.

Here, we demonstrate the microscale mapping of coupling strengths between resonator structures and metasurfaces and sub-wavelength localized THz dipoles on spintronic emitters. Utilizing microscale bow-tie geometries excited by 16-µm-sized optical foci, we observe pronounced changes in the emitted THz spectrum, providing an avenue for the tailoring of spectral and angular emission characteristics from spintronic THz emitters. Numerical simulations link these findings to antenna resonances mediating between the localized dipole and far-field radiation. For split-ring resonator metasurfaces, the shape of THz Bloch modes and edge states are extracted. These developments allow for an in-depth experimental tailoring and characterization of the THz photonic environment in close proximity to spintronic emitters.

## Results

For the generation of a micrometer-scale broadband THz emitter with a source size far below the THz diffraction limit, we utilize a resonator-grafted spintronic emitter bilayer (CoFeB(2 nm)/Pt(2 nm)) which is illuminated by a focused optical excitation pulse (70-fs FWHM pulse duration, 780 nm wavelength, 16-µm FWHM spot size), as depicted in Fig. 1a. The field polarization of the generated THz dipole is perpendicular to the local in-plane magnetization of the CoFeB layer and thereby tunable by the external magnetic field. The near field of the localized transient electric dipole is coupled to a metallic resonator structure in direct proximity to the emitter only separated by a $SiO_2$ (53-nm thickness) insulation layer[33]. For a varying position of the excitation focus, we detect the emitted THz time-domain signal in the far field utilizing an electro-optic sampling approach (see Methods for details).

**Local dipole coupled to a bow-tie resonator.** As a first example, we graft a gold bow-tie antenna (gap size: 10 µm) onto the spintronic emitter. Placing the THz dipole far away from the resonator structure, and with the CoFeB magnetization perpendicular to the gap (Fig. 1a), we observe a far-field transient (Fig. 1b, gray curve) with a dominant oscillation period of about 0.5 ps. In the frequency domain, the resulting THz spectrum (Fig. 1d, gray curve) shows a width of about 1.7 THz (FWHM), limited by the excitation pulse duration and the detection bandwidth, and a center-of-mass at 2.1 THz. Remarkably, the emitted THz signal changes drastically when the dipole is positioned within the gap of the bow tie (Fig. 1(b,d), blue curves). The emission spectrum now exhibits two maxima located at 1.2 THz and 2.55 THz, respectively, and an overall increased spectral bandwidth of 2.4 THz. Importantly, and different from the results by applying spectral filter[34], spectral amplitudes also exhibit an enhancement by about 55 % around a frequency of 1.2 THz, indicating pronounced resonances. Changing the in-plane magnetization direction (and thereby the THz field polarization) by 90° uncouples the dipole from the bow-tie resonator, and the reference THz spectrum observed without a grafted resonator is largely recovered (Fig. 1d, red curve). The differing coupling behavior for both polarizations is also found for the THz phase difference for emission from the bow-tie gap relative to reference spectra (Fig. 1d, lower panel).

To further extract the characteristic length scale of dipole-resonator coupling, we raster-scanned the optical excitation focus across the resonator. In Fig. 1c, the resulting overall THz spectral intensity (central panel) and the normalized ratio of the 1.2-THz and 2.55-THz spectral components (right panel) are shown. As a spatial reference, the simultaneously recorded optical transmission of the pump light is depicted in the left panel. The overall THz intensity is reduced by about one order of magnitude for the THz dipole located within the triangular gold region of the bow-tie resonator. Notably, the optical transmission in the gap almost equals the transmission in the reference region, ensuring that the excitation focus is not significantly clipped by the adjacent gold structure. Large spectral changes in the emitted THz radiation are only visible within a circular region in the gap within a diameter of about 20 µm, indicating the strongly localized interaction of the induced dipole with the surrounding metallic environment.

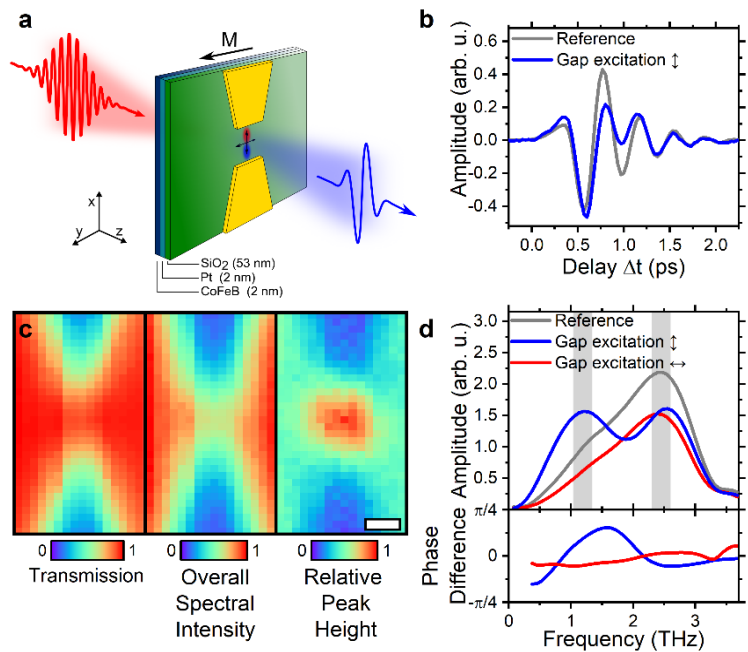

**Figure 1: Tailoring terahertz emission spectra by surface micro-structuring. a**, Experimental scheme. A localized THz electric dipole, driven by an optically triggered inverse spin Hall current, is coupled to the gap of a bow-tie resonator (sample magnetization: y-direction, THz polarization: x-direction). The emitted THz radiation is detected in the far field by electro-optic sampling. **b,d**, THz time-domain signal (**b**) and corresponding spectral intensity (**d**) obtained for a bow-tie-coupled THz dipole (blue curves) and for an unstructured emitter (gray curves). The pronounced changes in the emitted THz radiation for the bow-tie-coupled case vanish for the THz dipole polarization perpendicular to the bow-tie axis (red curve in **d**). The difference between the spectral phases for both polarizations with respect to the reference are shown in the lower panel. **c**, Spatial map of the optical transmission (left) and overall spectral intensity of the emitted THz radiation (center) were recorded simultaneously by raster scanning the optical excitation spot (spectral intensity normalized to maximum). The intensity ratio of the two spectral regions highlighted in **d** weighted by the overall spectral intensity is displayed in the right panel, visualizing the strong confinement of dipole-bow-tie coupling within the gap region (scale bar: 10 µm).

For investigating the spatial dependence of the emitted signal and its reproducibility, we patterned a series of gold bow-tie resonators onto the STE surface (optical micrograph in 2a) and recorded the THz emission for the optical focus being placed along a line through the resonator gaps (yellow dashed line). Figure 2(b,c) illustrate the obtained THz time-domain signal and THz spectrum as a function of the optical excitation focus position, exhibiting pronounced changes in the emitted THz radiation. In the spectra, the center-of-mass oscillates between lower and higher frequency components (gray curve). A selection of time traces and spectra are shown in 2(d,e) recorded with an excitation position ranging from in-gap (red) to in-between adjacent bow ties (blue). Notably, for an excitation in-between two bow ties, the total emitted intensity even exceeds the reference signal (intensity enhancement of about 23 %), indicating an antenna-like effect of the triangular side-edges of the resonator. Here, the overall temporal and spectral shape of the THz transient show only moderate changes compared to the reference pulse. Similar effects are observed in the vicinity of isolated bow-tie structures and at the out-most resonators in a linear array (see Fig. S6). Figure 2f shows the ratio of spectral amplitudes within the two gray marked areas in Fig. 2e as a function of the excitation position. The resulting feature widths of 19 µm (FWHM) demonstrates the pronounced local differences.

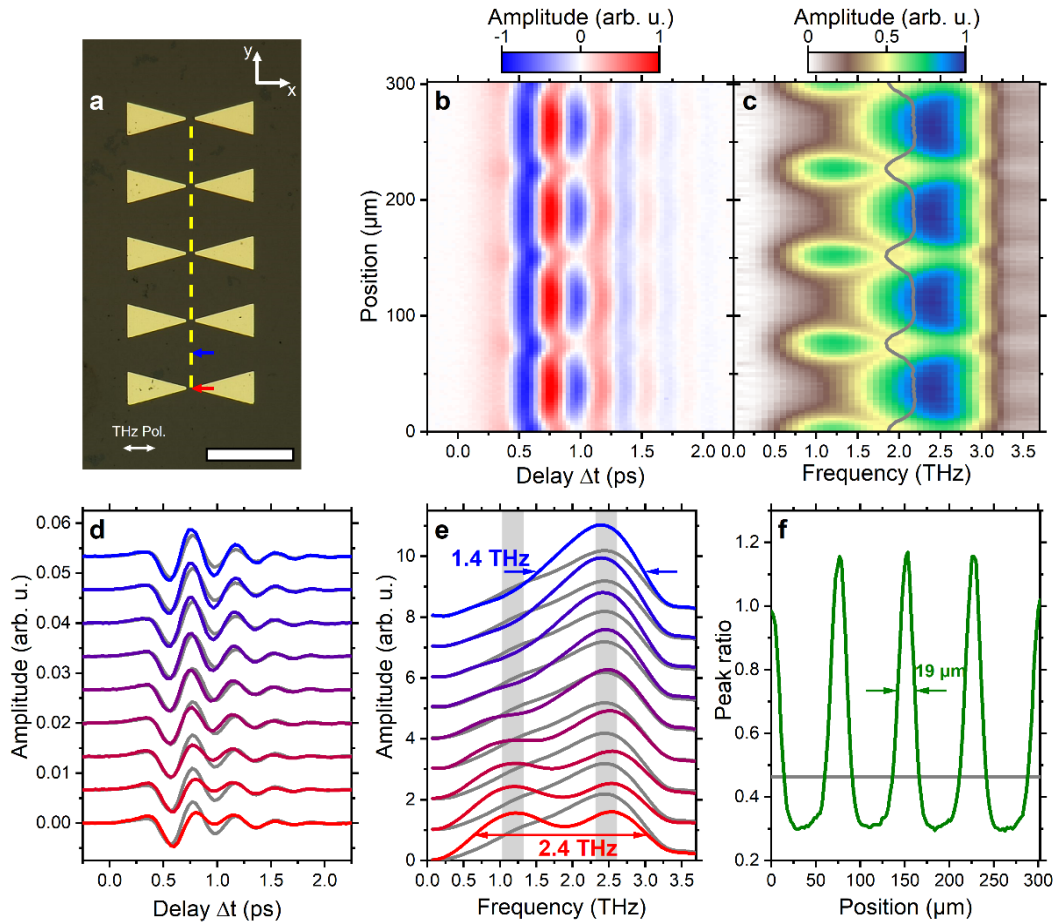

**Figure 2: Coupling of a THz dipole to a bow-tie resonator array. a,** Optical micrograph of gold bow-tie antennas deposited on a STE (gap size: 10 µm, THz polarization perpendicular to gap, scale bar: 100 µm). **b,c,** The position-dependent THz time-domain signal (**b**) and spectral intensity (**c**) recorded for an excitation pulse positioned along the yellow dashed line indicated in **a** shows prominent variations and an additional resonance peak for in-gap excitation. Each THz spectrum is normalized to its integrated intensity. Spectral center-of-mass frequency is indicated by gray curve. **d,e,** The exemplary selected time traces and (non-normalized) spectra for electric dipole positions changing from in-gap (red curve; position indicated in **a**) to a point between two adjacent resonator structures (blue, cf. **a**) (gray curves: THz reference signal for unstructured emitter; spectral widths indicate FWHM). **f,** The extracted ratio of the two spectral peaks marked in gray in **e** shows a distinct spectral shift towards the lower frequency peak for dipole positions in the resonator gap (local confinement width of 19 µm FWHM).

**Numerical model of dipole-resonator coupling.** In order to develop a theoretical framework of the local dipole-resonator coupling, we conducted finite-difference time-domain (FDTD) simulations (ANSYS Lumerical FDTD module) of a single bow-tie antenna excited by a dipole centered in the bow-tie gap (see Methods for details). The time-dependent dipole moment is chosen to be proportional to the experimental time-domain THz signal recorded on an unstructured surface (cf. Figure 1b, gray curve). Figure 3a displays the radiated electric field $E_x$ in a plane with a distance of 50 µm to the substrate for different delay times Δt. At early times (Δt = 0.92 ps), the electric field distribution shows an approximate circular symmetric phase as expected for the radiation pattern of an uncoupled dipole. At later times, a pronounced symmetry breaking in the radiation field can be observed, with the main axis linked to the orientation of the bow-tie antenna. From the field distribution close to the sample, we calculated the far-field emission pattern at different frequencies, as displayed in Fig. 3b. The angular emission pattern exhibits distinct changes with frequency, gradually evolving from a rather homogeneous distribution (slightly elongated along the $k_y$-direction) at 1.32 THz to distributions with two and three lobes at 2.42 THz and 2.82 THz, respectively (see also Fig. S4). As a comparison, we also calculated the far-field emission pattern of a dipole on a bare substrate, which is close to circular symmetric and shows no significant frequency dependence. Notably, the presence of a substrate already substantially modifies the far-field emission pattern as compared to a dipole in free space[47]. To obtain a microscopic understanding of the frequency-dependent angular emission patterns, Fig. 3c provides the calculated electric near-field phase in the plane of the bow-tie resonator. At 1.32 THz the electric field in the bow-tie gap oscillates in-phase with the field at the base of the bow-tie triangles, resulting in a constructive interference in the direction perpendicular to the surface. At 2.42 THz the field in the gap and base regions oscillate with a $\frac{4}{3}\pi$ phase difference so that the emission in the perpendicular direction interferes mainly destructively. At a frequency of 2.82 THz the different antenna parts oscillate with a phase difference of $\pi$ which yields a more complex angular far-field distribution. In the experiment, THz radiation is collected with a solid angle of 0.66 sr, corresponding to an opening angle of ±26.5° as defined by the first off-axis parabolic mirror after the sample. To account for the angular selectivity of the setup, we only consider the central region in the far-field pattern indicated by dashed circles in Fig. 3b. For the bow-tie-coupled dipole, the extracted overall spectral intensity in the central angular region (Fig. 3d, blue curve) features two pronounced peaks at 1.32 THz and 2.82 THz and a minimum at 2.42 THz, reproducing the main features in the experimental spectrum (Fig. 1d). Enhanced spectral intensities occur at frequencies with the angular emission distribution becoming localized around k = 0, yielding a more collimated radiation emitted perpendicular to the substrate. In contrast, at frequencies of about 2.42 THz the radiation is preferentially emitted with larger $k_x$-components not detected in the experiment. In contrast, the corresponding spectral intensity extracted for the bare substrate (Fig. 3d, gray curve) largely resembles the dipole excitation spectrum. Equally, the change in the spectral phase as observed experimentally (Fig. 1d, lower panel) is qualitatively reproduced in the simulation which, however, exhibits an overall larger spectral phase shifts and steeper phase gradients (Fig. 3d, green curve).

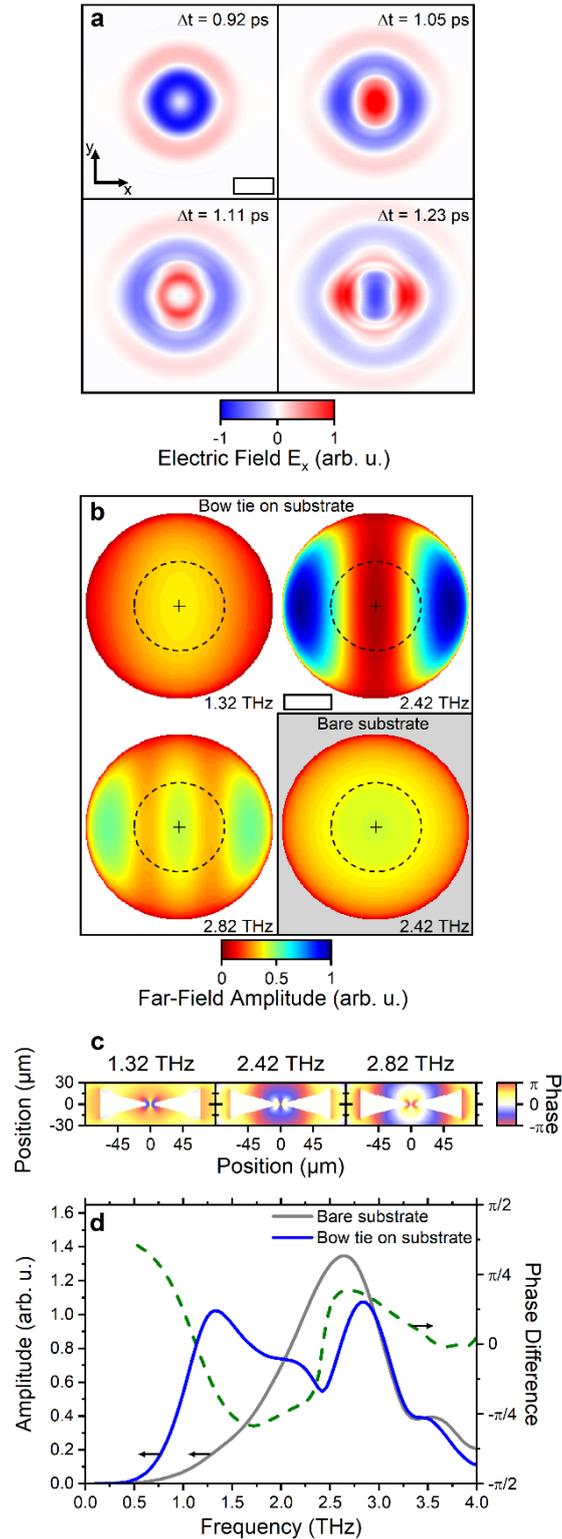

**Figure 3: Finite-difference time-domain calculation of bow-tie-coupled dipole radiation.** **a**, Radiated electric field $E_x$ for a localized dipole in the bow-tie gap evaluated at a 50-μm distant plane at different delay times (scale bar: 100 μm). **b**, Extracted angular far-field emission pattern for three exemplary frequencies. As a reference, the corresponding pattern for the dipole above a bare substrate is given in the lower right panel (emission direction perpendicular to the substrate is indicated by cross mark; scale bar: $\frac{1}{2}\sqrt{k_x^2 + k_y^2}/k$; dotted circle: acceptance angle of the experimental detection unit). **c**, The calculated near-field phase

for the three exemplary frequencies in **b** provide a microscopic mechanism for the observed frequency-dependent angular emission changes (see text for details). **d**, Simulated spectral intensity for a dipole in the bow-tie gap (blue curve) and above a bare substrate (gray curve), obtained by integrating the angularly resolved emission intensities in **b** within the acceptance angle. Green dashed curve: difference in spectral phase for both cases.

**THz mode mapping.** Finally, we demonstrate that the coupling of a THz dipole with resonator structures can be also investigated in the case of more complex surface patterns. For this purpose, a periodic array of gold split-ring resonators (SRR)[48] was deposited onto the spintronic emitter, as depicted in Fig. 4. Similar to the case of bow-tie resonators, the emitted THz spectrum depends on the excitation position, as exemplified in Fig. 4b for the three positions marked in Fig. 4a. For isolated split-ring resonators only weak dipole couplings are found (cf. Fig. S10 and S11). In order to illustrate the local emission changes, we performed an excitation raster scan within one unit cell of the array (indicated as square in Fig. 4a) and decomposed the recorded spectra by a principle component analysis (PCA)[49]. The three dominant extracted PCA components (A) to (C), as well as a spatial map of the associated coefficients are depicted in Figure 4c and 4d, respectively. The PCA component (A) largely mirrors the reference spectrum and has high negative coefficients at regions with an overall attenuated emission, for example at the gold-covered sample regions. The most distinct spectral change is detected close to the gap in the split-ring resonators, at which large coefficients for PCA component (B) are obtained. In particular, the exemplary spectrum exhibits an additional peak around 1.35 THz which corresponds to the dispersive line-shape in the PCA component (B). The expected size dependence of the resonance in the (B) component is experimentally verified by investigating THz emission from re-scaled SRR arrays with unit cells ranging from 40 µm to 96 µm (Fig. 4e). A spectrally and spatially distinct mode (component (C)) is located predominantly between SRRs in the vertical direction. Generally speaking, the extracted PCA components are expected to be related to photonic Bloch modes[50] in the periodic SRR array and their coupling to far-field radiation within the acceptance angle of the THz detection setup. A single THz resonance in a similar structure was recently identified by utilizing scattering SNOM with narrow bandwidth excitation[51].

To highlight the difference between THz modes within a periodic array and at the array edges, we performed a line scan of the optical excitation spot across a 6x6 SRR array, as indicated by the yellow dashed line in Fig. 4a. The corresponding THz spectra are given in Fig. 4f. Within the array, the spectral changes as extracted from PCA are recovered. In addition, at the positions smaller zero or larger than 480 µm edge states are apparent. Notably, the low symmetry of the SRR unit cell along the scan direction is also evident in a spatial shift of the spectral intensity minimum at 1.8 THz by about 7 µm towards the gap side of the split-ring resonator, resulting in staggered configuration and a slight difference between the left and right edge state. In future studies, the relationship between properties of the emitted THz radiation and the topological character of the photonic edge state[52,53] could be of particular interest.

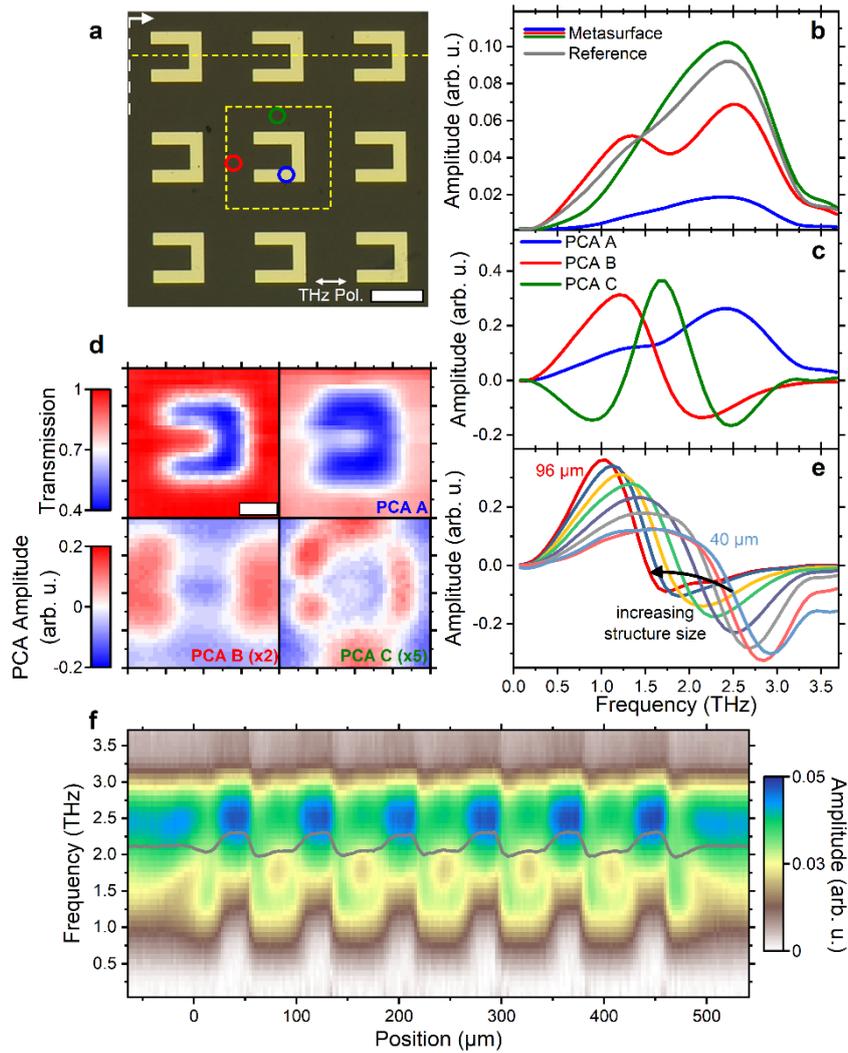

**Figure 4: Mapping optical modes in metasurfaces by localized THz dipole excitation. a**, Optical micrograph of a two-dimensional gold split-ring resonator (SRR) array deposited onto a spintronic THz emitter (scale bar: 40 µm). **b**, Exemplary THz spectra for the THz dipole placed at different positions in the unit cell of the metasurface. Line colors correspond to the locations indicated by open circles in **a**. Gray curve: reference THz spectrum from an unstructured emitter. **c,d**, Raster scanning of the optical excitation pulse position across an SRR unit cell, in the region indicated by the yellow dashed square in **a**, yields a transmission map (**d**, upper left) and THz time-domain signals for each position (scale bar: 20 µm). A principle component analysis (PCA) decomposes the resulting spectra into three dominant spectral components (**c**) with corresponding coefficient maps (A) (**d**, top right), (B) (**d**, bottom left) and (C) (**d**, bottom right). **e**, PCA spectral component (B) extracted for scaled metasurfaces with unit cells ranging from 40 µm to 96 µm yield a frequency-dependent shift of the maximum of PCA component. **f**, THz spectra recorded along a line through an array of six resonator structures (scanning position (yellow dashed line) and zero-position (white dashed arrow) at left edge of first SRR unit cell both indicated in **a**) shows significant spectral variations within the array and the presence of edge modes. Each spectrum is normalized to its integrated intensity. Gray curve: spectral center-of-mass.

## Conclusion and Outlook

In conclusion, we have investigated the emission properties of locally excited micro-patterned STE. By grafting metallic structures onto the emitter surface, we demonstrated flexible tailoring of spectral emission properties, exhibiting a significant spectral enhancement at distinct frequency components. Such a compact tailored on-chip THz spintronic device shows a tuneable spectral bandwidth and angular emission pattern. Employing localized THz dipole excitation allows for a direct coupling to THz near fields which we utilized for the mapping of THz modes in metasurfaces. Potential applications include the characterization of topological THz systems as well as the generation of high field strengths in microresonators.

## Methods

**Experimental setup.** The experiments were carried out in an optical THz spectroscopy setup (cf. Figure S1a). Short infrared laser pulses from a frequency doubled erbium-doped femtosecond fiber laser (pulse duration: 70 fs (FWHM), wavelength: 780 nm; repetition rate: 100 MHz; C-Fiber 780 from Menlo Systems) were focused onto the structured spintronic terahertz emitter sample (16-μm-FWHM focal spot size, pulse energy: 0.4 nJ) under dry-nitrogen atmosphere. The external magnetic field of a pair of neodymium magnets saturates the in-plane magnetization of the STE. The THz transient was detected in an electro-optic sampling setup[54,55]. Here, the THz pulse induces birefringence in a 500-μm thick <110>-oriented ZnTe electro-optic crystal proportional to the instantaneous electric field, rotating the polarization of a probing pulse. Its polarization state for different time delays of the THz and probing pulse was recorded behind a Wollaston prism using balanced photodiodes. For obtaining a higher signal-to-noise ratio, a lock-in amplification was applied with an optical chopper wheel installed in the excitation pulse arm (modulation frequency: 140 Hz). For automated raster scanning, the emitter position was controlled by a three-axis piezo-motor-driven translation stage system.

The detection angle sensitivity of our setup is limited to approximately ±26.5° due to the collection angle of the off-axis parabolic mirror located after the spintronic terahertz emitter (Fig. S1b). A silicon wafer tilted to the THz Brewster angle transmits the p-polarized THz transient and reflects the residual 780-nm beam onto a photodiode detector. In this configuration, we recorded the optical transmission at each excitation pulse position simultaneously to performing THz time-domain scans.

For the experiments reported here, the magnetic field was oriented in the vertical direction, to maintain p-polarized THz emission. The relative orientation between the THz polarization direction and the microstructures was adjusted by rotation of the emitter.

For obtaining the emitted THz spectra, acquired delay-dependent EOS-signals are filtered by a super-Gaussian window function (width: 2.5 ps FWHM) to exclude trailing replica pulses (resulting from reflections on optical components) from the analysis. In addition, we applied zero padding of the time-domain signal before performing a Fourier transform of the signal.

**Sample preparation.** We utilize a spintronic terahertz emitter consisting of a 2-nm thin ferromagnetic CoFeB layer and a 2-nm thin non-magnetic Pt layer on fused silica glass. The CoFeB layer is magnetron sputtered onto the substrate, directly followed by the Pt layer being deposited by e-beam evaporation without breaking the vacuum conditions in a separate ultrahigh-vacuum chamber with a base pressure of $5 \times 10^{10}$ mbar. To avoid direct injection of electron currents from the spintronic bilayer into the microresonator structures, the bilayer was capped by an insulating $SiO_2$ layer (thickness: 53 nm).

For structuring the spintronic emitter, electron-beam lithography (EBL) was employed. A double-layer positive resist film (PMMA E-Beam Resist AR-P 632.06 (50K) and AR-P 672.03 (950K) from Allresist) was subsequently spin-coated (2000 revolutions per minute, with 3 minutes of heating at 150°C) onto the emitter. Electron beam exposure was carried out in a Helios NanoLab 600i (FEI) equipped with an EBL module (ELPHY-Quantum (Raith)) at a dose per area of about 150 µC/cm$^2$ and an electron kinetic energy of 20 keV and beam current of 0.34 nA.

The exposed sample was developed for 1 min (E-Beam Developer AR 600-55 from Allresist) and transferred into an electron-beam evaporator (Pfeiffer Vacuum PLS 500) where 35 nm gold were deposited with a 5-nm chromium adhesion layer. Subsequently, the lift-off of residual resist was performed in an acetone bath resulting in isolated gold microresonators grafted on the emitter surface.

**Resonator geometries.** In our studies, we examined three different gold bow-tie antenna designs deposited onto the spintronic THz emitter surface, with the sizes scaled evenly (cf. Fig. S3a). In addition to the structures mentioned in the main text[56] (length $l_1$ = 140 µm, height $h_1$ = 37.5 µm, gap size $g_1$ = 10 µm), bow-tie structures with geometrical dimension of $l_2$ = 70 µm, $h_2$ = 18.75 µm, $g_2$ = 5 µm and $l_3$ = 210 µm, $h_3$ = 56.3 µm, $g_3$ = 15 µm were investigated. The split-ring resonators mapped in Fig. 4 of the main text have the dimensions of gap size $g$ = 20 µm, wire width $w$ = 10 µm, length $l$ = 40 µm and unit cell size = 80 µm (cf. Fig. S3b). Additionally, the size of the split-ring resonator arrays was varied from $l$ = 20 µm to 48 µm in 4-µm increments ($g$ = $l$/2, $w$ = $l$/4 and unit cell = 2$l$).

**Numerical simulations.** To obtain a microscopic understanding of the dipole-resonator coupling, we carried out numerical finite-difference time-domain simulations of the electric dipole coupling to a bow-tie antenna using a commercial software package (ANSYS Lumerical FDTD module). In a three-dimensional simulation geometry, a gold bow-tie structure (thickness: 35 nm) is placed on top of a SiO$_2$ substrate (thickness: 100 µm). We applied the Drude model to approximate the dielectric properties of gold at terahertz frequencies (plasma frequency: $1.38 \cdot 10^{16}$ rad/s; permittivity: 1, plasma collision frequency: $4 \cdot 10^{13}$ rad/s[57]). The experimentally recorded time-domain signal (cf. Figure 1b) is used as a source signal for an electric point-dipole positioned in the bow-tie gap near the substrate-air interface (10 nm displaced from the interface, polarization in x-direction). At the simulation box boundaries, perfectly matched layers are applied to avoid back reflections of the electromagnetic waves. The temporal and spatial THz field distribution is recorded in a plane just above the dipole, as well as in a plane at z = 50 µm. After applying an apodization to exclude residual back reflections from the simulation boundaries, the spectral THz components are extracted using a temporal Fourier transformation.

For the calculation of the frequency-dependent far field[58], the recorded field distribution in the 50-µm plane was used. A spatial window function was employed to avoid electric field truncation at the edges of the sampling plane (far-field filter setting: α = 0.05). The phase of the electric near-field was extracted from a temporal Fourier transformation of the electric field calculated in the bow-tie antenna plane. To obtain the overall spectral intensity as depicted in Fig. 3d, we only considered the central portion of the far field with an effective radius of ±26.5°, equal to experimental THz detection angle.

## Data availability

The data that support the plots within this paper and other findings of this study are available from the corresponding authors on reasonable request.

## Acknowledgements

We acknowledge financial support by the Volkswagen Foundation as part of the Lichtenberg Professorship "Ultrafast nanoscale dynamics probed by time-resolved electron imaging". We thank the research group of C. Lienau (University of Oldenburg) for the possibility of temporarily using their femtosecond laser system in an early phase of the project. We also thank H. Koch and M. Macke for technical support. The authors thank S.-A. Biehs for helpful discussions.

## Author contributions

C.R. and R.v.S. designed the experiment, performed the measurements and analyzed the data with contributions from S.S.. The spintronic THz emitters were fabricated by J.K., J.W. and M.M.. Grafting of the spintronic THz emitter was done by C.R. with contributions from L.A.G.. The FDTD simulations were carried out by C.R.. The manuscript was written by C.R., R.v.S. and S.S. after discussion with and input from all authors.

## Competing interests

The authors declare no competing interests.

# Supplement

Experimental Methods

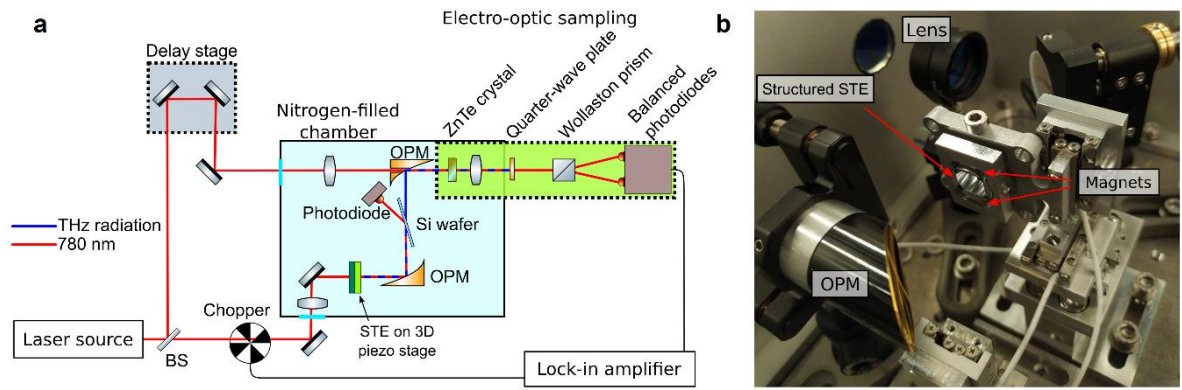

**Figure S1: Experimental THz time-domain spectroscopy setup. a**, Sketch of the optical setup. Focused femtosecond optical pulses excite a structured spintronic terahertz emitter (STE). The THz radiation is collected and detected by electro-optic sampling in the far field. BS: Beam splitter, OPM: Off-axis parabolic mirror. **b**, Photograph of the STE positioning stage and the THz beam-defining off-axis parabolic mirror.

Homogeneity of THz emission and acquisition of reference scans

For the reference signals shown in the main text we averaged over 40 transients, each acquired at a different excitation pulse position in an unstructured area of the spintronic emitter. Figure S2 depicts THz spectra recorded along a line, which exhibit no significant variation of the spectral shape and center-of-mass (black curve).

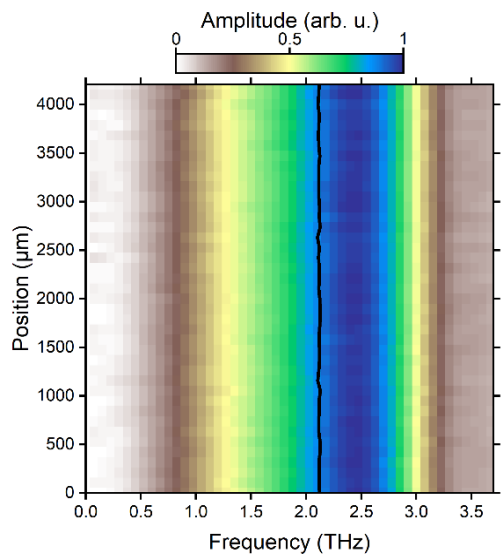

**Figure S2: The position-dependent THz spectral intensity recorded for an excitation pulse positioned along a line in an unstructured area of the spintronic emitter.** Each THz spectrum is normalized to its integrated intensity. Spectral center-of-mass is indicated by the black curve.

Resonator geometries

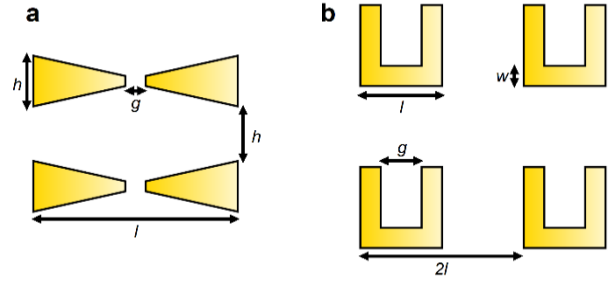

**Figure S3: Schematic representation of the resonator designs. a**, Bow-tie antenna with length *l*, height *h* and gap size *g*. The distance of adjacent structures equals *h*. **b**, The split-ring resonators with a length and height of *l* feature a gap size of *d* = *l*/2 and a wire width of *w* = *l*/4 (unit cell: 2*l*).

Numerical simulations

We further compared the far-field radiation characteristics for different bow-tie resonator sizes and dipole polarizations (cf. Figure S4 and S5). A distinct angular dependence is observed for a dipole polarization perpendicular to the resonator gap (Fig. S4(a,c)), whereas no evident change is noticeable for a parallel dipole orientation (Fig. S4b). As a reference, we also calculated the spectrum of a dipole on bare substrate exhibiting no spectral modulation and angular dependence (Fig. S4d). In addition, Figure S5 depicts the extracted measured and simulated spectral intensities for a dipole placed in the gap of a bow tie with dimensions scaled by a factor of 1.5 relative to the resonator geometry of Fig. 1 and 2. Here, the spectral features in the experimental results due to antenna coupling are well reproduced by the numerical simulation.

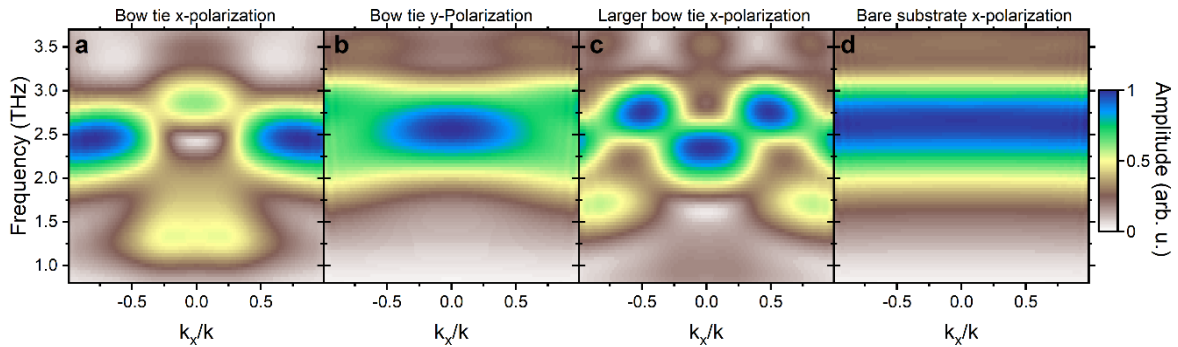

**Figure S4: Extraction of the simulated far-field distribution along $k_x$-direction ($k_y$ = 0) for a dipole coupled to a bow-tie resonator gap for different emission frequencies.** The dipole oscillates in x- (**a**) and y-direction (**b**) for a resonator size of *l* = 140 µm, and *l* = 210 µm (**c**, x-polarization). While a pronounced spectral change of the far-field pattern is evident for the dipole oscillating perpendicular to the resonator gap (**a,c**), only a moderate effect can be observed for the dipole polarized along y-direction. Moreover, no angle dependence of the far-field emission spectrum is apparent for the dipole on bare substrate (**d**). Each THz spectrum is normalized to its integrated intensity. (experimental detection angle: $\frac{k_x}{k} = \pm 0.46$).

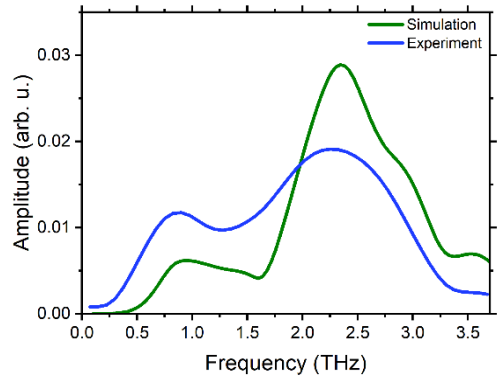

**Figure S5: Experimental and calculated far-field spectrum for a dipole coupled to a larger bow-tie structure.** Although the THz near-field emission is different for the experiment and FDTD simulation, the spectral modulation can be reproduced. The bow-tie dimensions are scaled by a factor of 1.5 compared to the resonator geometry of Fig. 1 and 2. Each THz spectrum is normalized to its integrated intensity.

Size dependence and array effects for THz dipole coupling to bow-tie resonator

In addition to the bow-tie array studied in the main text, we investigated the THz emission spectra for an isolated bow tie (Fig. S6a) and the lowest resonator of a series of bow ties (Fig. S6b). For both cases, the emitted THz spectrum for gap excitation is similar to the case of bow-tie arrays. Also, for out-of-gap excitation close to the gap (at a distance of about 32 µm) a distinct emission enhancement is observed.

Figure S7 depicts the spectral dependence on the optical excitation pulse position for bow-tie arrays with structures scaled by a factor of 0.5 and 1.5 relative to the dimensions of the resonator in Fig. 1 and 2. In the spectra from smaller and larger bow-ties, the maximum shifts from 0.9 THz for the largest structure to 2.25 THz for the smallest structure.

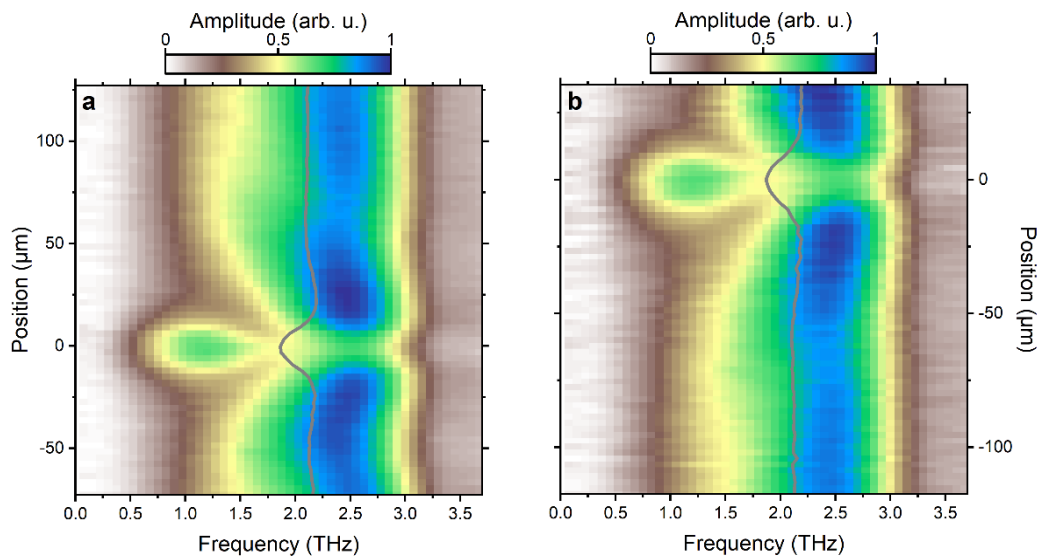

**Figure S6: The position-dependent THz spectral intensity recorded for an excitation pulse positioned along a line through the gaps of an isolated bow-tie structure (a) and the lowest resonator of a series of bow ties (b).** The bow-tie dimensions equal the resonator geometry of Fig. 1 and 2. The zero-position corresponds to the bow-tie gap location. Each THz spectrum is normalized to its integrated intensity. Spectral center-of-mass is indicated by gray curve. THz polarization is perpendicular to gap orientation.

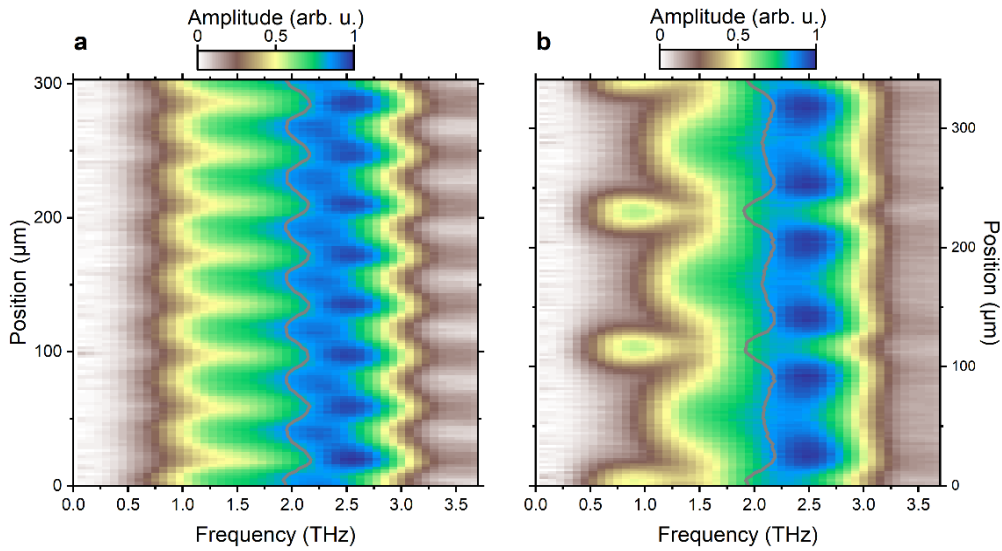

**Figure S7: The position-dependent THz spectral intensity recorded for an excitation pulse positioned along a line through the gaps of a one-dimensional bow-tie resonator array.** (THz polarization perpendicular to gap orientation). The bow-tie length amounts to 70 μm (**a**, 1x9-array) and 210 μm (**b**, 1x4-array). Each THz spectrum is normalized to its integrated intensity. Spectral center-of-mass is indicated by gray curve.

THz mode mapping in different metasurfaces

To complement the investigation of THz emission from split-ring resonator arrays in Fig. 4, we also prepared emitter with a closed-gap geometry (Fig. S8) and with the gap orientation parallel to the THz polarization (Fig. S9). In Fig. S8, similar mode patterns as in Fig. 4 are observed, except that the slight symmetry breaking seen for the split-ring resonator modes is no longer visible. Changing the relative polarization of the THz radiation with respect to the structure (Fig. S9) results in distinct differences of the modal patterns. Furthermore, Fig. S10 displays the PCA analysis for an isolated SRR. For a comparison of the spectral effect of different resonator configurations, we depicted example spectra in Figure S11. Here, the optical excitation pulse position corresponds to the unit cell location indicated by open red circle in Fig. 4a.

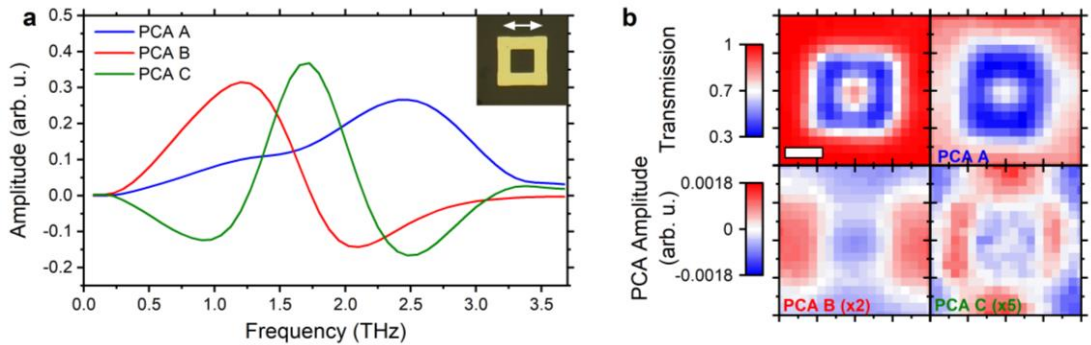

**Figure S8: Principle component analysis for a closed split-ring resonator structure within a 5x5 array.** (inset: optical micrograph of a single unit cell, resonator length: 40 μm, white arrow indicates THz polarization direction). **a**, The three dominant PCA components with corresponding coefficient maps (A) (**b**, top right), (B) (**b**, bottom left) and (C) (**b**, bottom right)

are extracted from a full unit cell raster scan of the optical excitation pulse as explained in the main text (scale bar: 20 µm). A transmission map (**b**, upper left) was subsequently recorded.

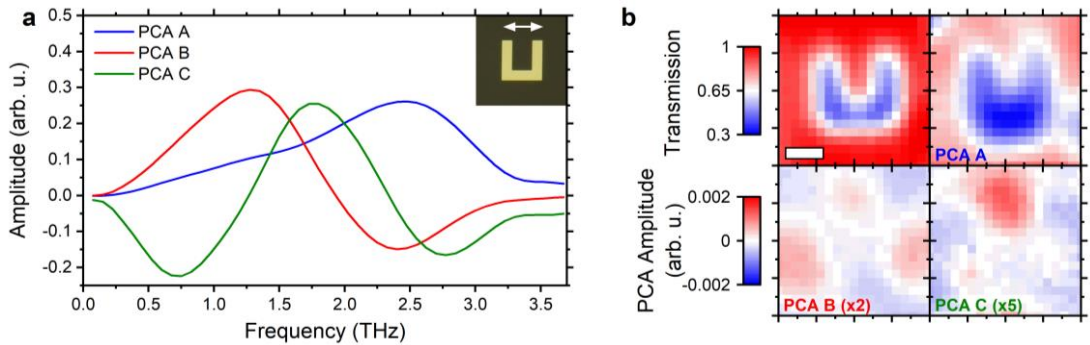

**Figure S9: Principle component analysis for a split-ring resonator structure within a 6x6 array with gap-orientation rotated by 90° compared to Fig. 4.** (inset: optical micrograph of a single unit cell, resonator length: 40 µm, white arrow indicates THz polarization direction). **a**, The three dominant PCA components with corresponding coefficient maps (A) (**b**, top right), (B) (**b**, bottom left) and (C) (**b**, bottom right) are extracted from a full unit cell raster scan of the optical excitation pulse as explained in the main text (scale bar: 20 µm). A transmission map (**b**, upper left) was subsequently recorded.

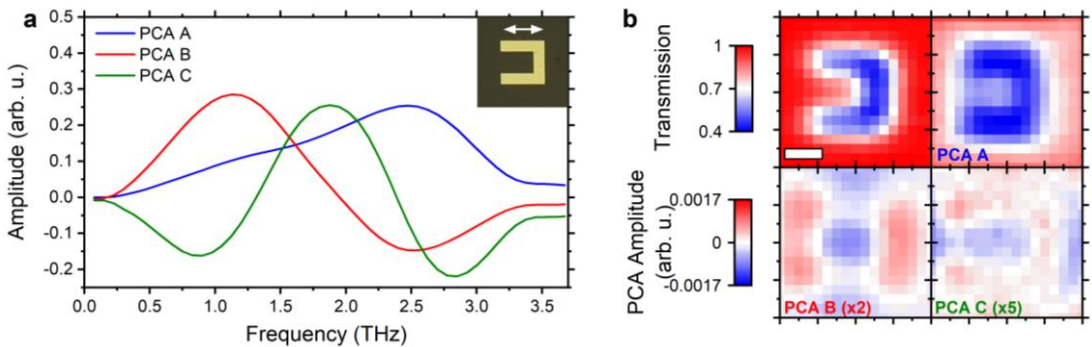

**Figure S10: Principle component analysis for an isolated split-ring resonator structure.** (inset: optical micrograph of a single unit cell, resonator length: 40 µm, white arrow indicates THz polarization direction). **a**, The three dominant PCA components with corresponding coefficient maps (A) (**b**, top right), (B) (**b**, bottom left) and (C) (**b**, bottom right) are extracted from a full unit cell raster scan of the optical excitation pulse as explained in the main text (scale bar: 20 µm). A transmission map (**b**, upper left) was subsequently recorded.

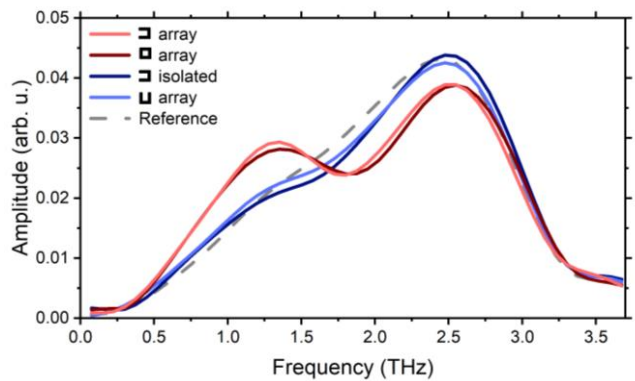

**Figure S11: Experimental spectra for different resonator arrangements.** The THz spectra for the resonator configurations shown in Fig. S8-S10 were extracted for dipoles placed at unit cell positions indicated by the red open circle in Fig. 4a. Distinct spectral changes, comparable to the SRR array from Fig. 4 (light red curve), are observed for an array of closed SRR structures (dark red curve). The spectra from the isolated resonator structure (dark blue curve) and an array of rotated SRR (light blue curve) show no significant difference to the reference spectrum (gray curve). Each spectrum is normalized to its integrated intensity.